\documentclass[aps, pre,showpacs,twocolumn,showemail]{revtex4}
\usepackage{bbm}
\usepackage{mathrsfs}
\usepackage[sans]{dsfont}
\usepackage{epsfig,psfrag}
\usepackage{amsmath,amsfonts,amssymb}
\usepackage[usenames]{color}
\usepackage[dvips, bookmarks, colorlinks=true, plainpages = false, citecolor = blue, linkcolor = blue, urlcolor = blue, filecolor = blue]{hyperref}
\def\be{\begin{equation}}
\def\ee{\end{equation}}
\def\bea{\begin{eqnarray}}
\def\eea{\end{eqnarray}}

\begin{document}

\preprint{draft}

\title{Fractal structure of a three dimensional Brownian motion on an attractive plane}
\author{Abbas Ali Saberi}\email{a$_$saberi@ipm.ir;  ab.saberi@gmail.com}
\address {School of Physics, Institute for Research in Fundamental
Sciences (IPM), P.O.Box 19395-5531, Tehran, Iran \\ Institut f\"ur
Theoretische Physik, Universit\"at zu K\"oln, Z\"ulpicher Str. 77,
50937 K\"oln, Germany}
\date{\today}

\begin{abstract}
Consider a Brownian particle in three dimensions which is attracted
by a plane with a strength proportional to some dimensionless
parameter $\alpha$. We investigate the fractal spatial structure of
the visited lattice sites in a cubic lattice by the particle around
and on the attractive plane. We compute the fractal dimensions of
the set of visited sites both in three dimensions and on the
attractive plane, as a function of the strength of attraction
$\alpha$. We also investigate the scaling properties of the size
distribution of the clusters of nearest-neighbor visited sites on
the attractive plane, and compute the corresponding scaling exponent
$\tau$ as a function of $\alpha$. The fractal dimension of the
curves surrounding the clusters is also computed for different
values of $\alpha$, which, in the limit $\alpha\rightarrow\infty$,
tends to that of the outer perimeter of planar Brownian motion i.e.,
the self-avoiding random walk (SAW). We find that all measured
exponents depend significantly on the strength of attraction.
\end{abstract}


\maketitle
\section{Introduction}

The laws of Brownian motion, formulated first by Einstein more than
a century ago \cite{Einstein}, have now found so many applications
and generalizations in all quantitative sciences \cite{Haw}. Many
fractal structures in the nature can be derived from the sample
paths of Brownian motion characterized by some appropriate fractal
dimensions \cite{Mandelbrot}.

A $d$-dimensional Brownian motion is known to be recurrent, i.e.,
the particle returns to the origin, for $d\leq$2 and escapes to
infinity for $d>$2. It is also known that the fractal (Hausdorff)
dimension of the graph of a Brownian motion is equal to 3/2 for
$d=$1, and 2 for $d\geq$2.

The scaling limit of interfaces in various critical 2$d$ lattice
models are proven or conjectured to be described by the family of
conformally invariant random curves i.e., Schramm-Loewner evolution
(or SLE$_\kappa$) \cite{schramm} which is driven by a 1$d$ Brownian
motion of diffusivity $\kappa$ \cite{SLE}.

One of the most important invariance properties of planar Brownian
motion is conformal invariance. Although the scaling limit of 2$d$
random walk, i.e., 2$d$ Brownian motion, because of self-crossing
itself does not fall in the SLE category, variations of Brownian
motion are described by SLE. Loop erased random walk (LERW) where
loops are removed along the way, is one of the examples which has
been studied by Schramm and shown that can be described by
SLE$_2$.\\The external perimeter of 2$d$ random walk is also a
non-intersecting fractal curve which can be defined by SLE.
Verifying an earlier conjecture by Mandelbrot \cite{Mandelbrot}, it
has been proven using SLE techniques \cite{Lawler} that the fractal
dimension of the Brownian perimeter is $d_f=$4/3, i.e, the same as
the fractal dimension of self-avoiding random walk (SAW) and the
external perimeter of the percolation hull.

In this paper, we investigate the statistical and fractal properties
of a 3$d$ random walker which is attracted by a plane. We believe
that this study can provide useful intuitive extensions for many
related physical phenomena including the problems with a discrete
time lattice walk \cite{appl0, appl1}, relaxation phenomena
\cite{relax}, exciton trapping \cite{trap} and diffusion-limited
reactions \cite{appl1, react}.

\section{The model}

We consider a random walker moving along the bonds of a cubic
lattice with the \emph{xy}-plane as an attractive plane. The
'walker' source is considered to be the origin of the coordinate
system. At each lattice point with $z\neq0$, there are six
possibilities for the random walker to select a link and move along.
In our model, the random walker prefers walking on and near the
attractive plane, and thus the probability that the random walker
chooses the link which approximates it to the attractive plane is
set to be $\alpha p$, and for remaining five links is considered to
be $p$, such that $\alpha>1$ (and will be called \emph{the strength
of attraction}) and $p=\frac{1}{\alpha+5}$. For each lattice point
on the attractive plane with $z=0$, the probability that each of the
four links on the plane to be chosen is set to be $\alpha p'$ and
for two other links perpendicular to the plane is considered to be
$p'$, where $p'=\frac{1}{4\alpha+2}$. The single parameter $\alpha$
in our model, controls the strength of attraction. Note that in the
limiting case $\alpha\rightarrow\infty$ our model reduces to the
pure 2$d$ random walk on the plane, and for $\alpha=1$ the pure 3$d$
random walk would be recovered.\\Thus there are four possible
probabilities. $\alpha p'$ for links that are in the attractive
plane, $p'$ for links from the attractive plane to either of the
neighboring planes, $p$ for links in all of the neighboring planes
or leading from them into the bulk, and $\alpha p$ for links from
all the neighboring planes to the attractive plane.\\By detailed
balance, in equilibrium at inverse temperature $\beta$, the ratio
$\alpha p / p'$ of the probabilities onto and off the attractive
plane, defines an attraction energy $\beta\epsilon = \ln[2\alpha
(1+2\alpha)/(\alpha +5)]$.

\section{Fractal dimension of the set of all visited sites and its level set}

%%%%%%%%%%%%%%%%%%%%%%%%%%%%%%%%%%%%%%%%%%%%%%%%%%%%%%%%%%%%%%%%%%%%%%%%%%%%%%%%%%%%%%%%%%%%%%%%%%%%%%%%%%%%%%%%
\begin{figure}[b]\begin{center}
\includegraphics[scale=0.39]{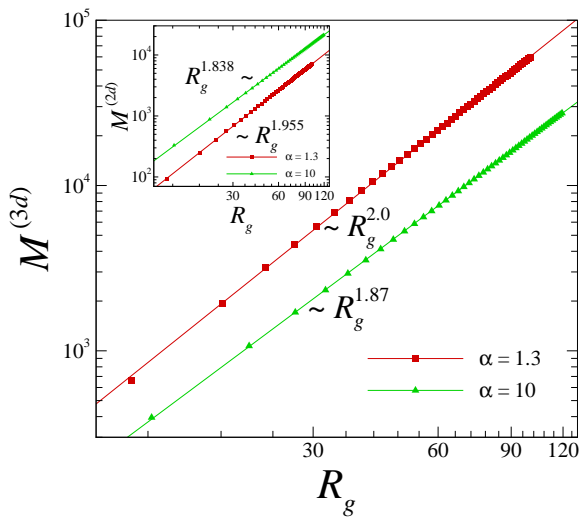}
\narrowtext\caption{\label{Fig0}(Color online) The average number of
total lattice sites $M^{(3d)}$ visited (at least) once by the
attracted random walker (ARW) (main panel), and those $M^{(2d)}$ on
the attractive plane (inset), as function of their average radius of
gyration for two different values of the strength of attraction
$\alpha=$1.3 ($\blacksquare$) and $\alpha=$10 ($\blacktriangle$).
The solid lines show the best fit to our data. The error bars are
almost the same size as the symbols.}\end{center}
\end{figure}
%%%%%%%%%%%%%%%%%%%%%%%%%%%%%%%%%%%%%%%%%%%%%%%%%%%%%%%%%%%%%%%%%%%%%%%%%%%%%%%%%%%%%%%%%%%%%%%%%%%%%%%%%%%%%%%%

\begin{figure}[t]\begin{center}
\includegraphics[scale=0.39]{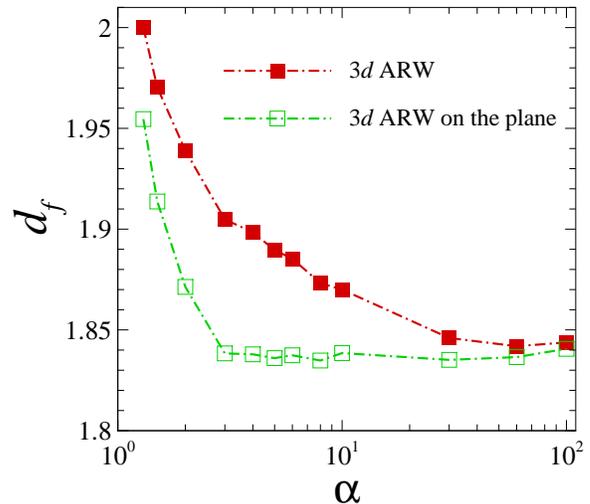}
\narrowtext\caption{\label{Fig1}(Color online) The fractal dimension
of the set of all lattice points visited (at least) once by the
attracted random walker (ARW) ($\blacksquare$), and the set of all
number of visited points on the attractive plane ($\square$), as
function of the strength of attraction $\alpha$. The error bars are
almost the same size as the symbols.}\end{center}
\end{figure}

In the cases of random walks, systems exhibit a \emph{generic scale
invariance}, meaning that the systems can exhibit self-similarity
and power laws without special tuning of parameters. This is why we
already expect that our model would exhibit rich fractal properties
for all values of $\alpha$.\\Let us first look at the fractal
spatial structure of the 3$d$ \emph{attracted} random walk (ARW) and
its intersection with the attractive plane. In order to estimate the
fractal dimension $d_f$ of the set of points visited (at least) once
by the random walker, we examine the scaling relation between the
average number of such points $M^{(3d)}$ and their corresponding
radius of gyration $R_g$, i.e., $M^{(3d)}\sim R_g^{d_f}$. Each
ensemble averaging for $M^{(3d)}$ (and also for $M^{(2d)}$ in the
following) and $R_g$ was taken over $5\times 10^4$ independent
samples for a fixed number of random walk steps $N$. The
measurements were done for $10^3\leq N\leq 10^5$ with the number
interval $\delta N=2\times 10^3$. We have also computed the fractal
dimension of the total number of sites on the attractive plane
(i.e., $M^{(2d)}$) visited by the random walker (in this case the
corresponding radius of gyration is computed for all set of distinct
visited sites only on the attractive plane $-$ see Fig.
\ref{Fig0}).\\ We find that the fractal dimensions have a remarkable
continuous dependence on the parameter $\alpha$. The results of
these fractal dimensions as function of the strength of attraction
$\alpha$ are illustrated in Fig. \ref{Fig1}. As can be seen from
figure \ref{Fig1}, for large values of $\alpha$, since the problem
reduces to the 2$d$ random walk on the attractive plane, these two
fractal dimensions converge to the same value close to the value
$\sim1.83$ (this is comparable with the fractal dimension of the set
of distinct sites visited by an 2$d$ RW on a square lattice, deduced
from the results reported in \cite{Lee}).
\\All error bars in this paper are estimated by using the standard
least-squares analysis, and are almost of the same size as the
symbols used in the figures.
\\For an ideal linearly self-similar fractal of dimension $d_f$, one
expects that the fractal dimension of the intersection being
$d'_f=d_f-1$ \cite{Mandelbrot}. But this is not apparently the case
here for $\alpha\neq 1$, since in our model, the attractive plane
has disturbed the homogeneity of the probability distribution in the
\emph{z}-direction. Only for $\alpha=1$ where $d_f=2$ \footnote{The
random walk on a simple cubic lattice is a \emph{transient} process,
since it has a finite escape probability $\approx$ 0.66. Therefore,
the number of distinct visited sites by the random walker is almost
the same as the number of steps or equivalently the trajectory
length, and thus, it is expected for both to have a same fractal
dimension 2.}, we find $d'_f=1=d_f-1$.
%%%%%%%%%%%%%%%%%%%%%%%%%%%%%%%%%%%%%%%%%%%%%%%%%%%%%%%%%%%%%%%%%%%%%%%%%%%%%%%%%%%%%%%%%%%%%%%%%%%%%%%%%%%%%%%%
%%%%%%%%%%%%%%%%%%%%%%%%%%%%%%%%%%%%%%%%%%%%%%%%%%%%%%%%%%%%%%%%%%%%%%%%%%%%%%%%%%%%%%%%%%%%%%%%%%%%%%%%%%%%%%%%

\section{Cluster size distribution on the attractive plane}

\begin{figure}[h]\begin{center}
\includegraphics[scale=0.28]{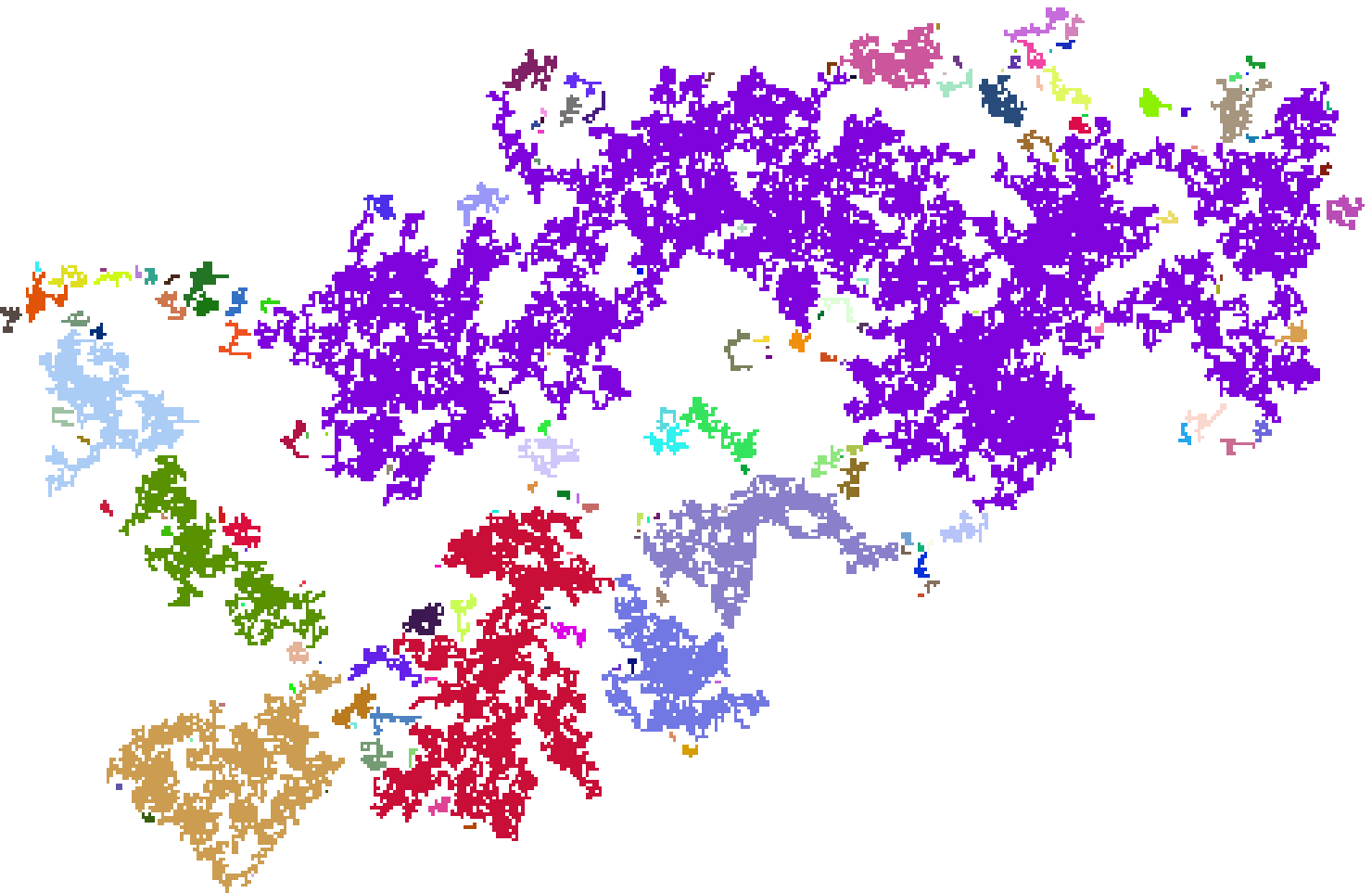}\hspace{0.5cm}\includegraphics[scale=0.23]{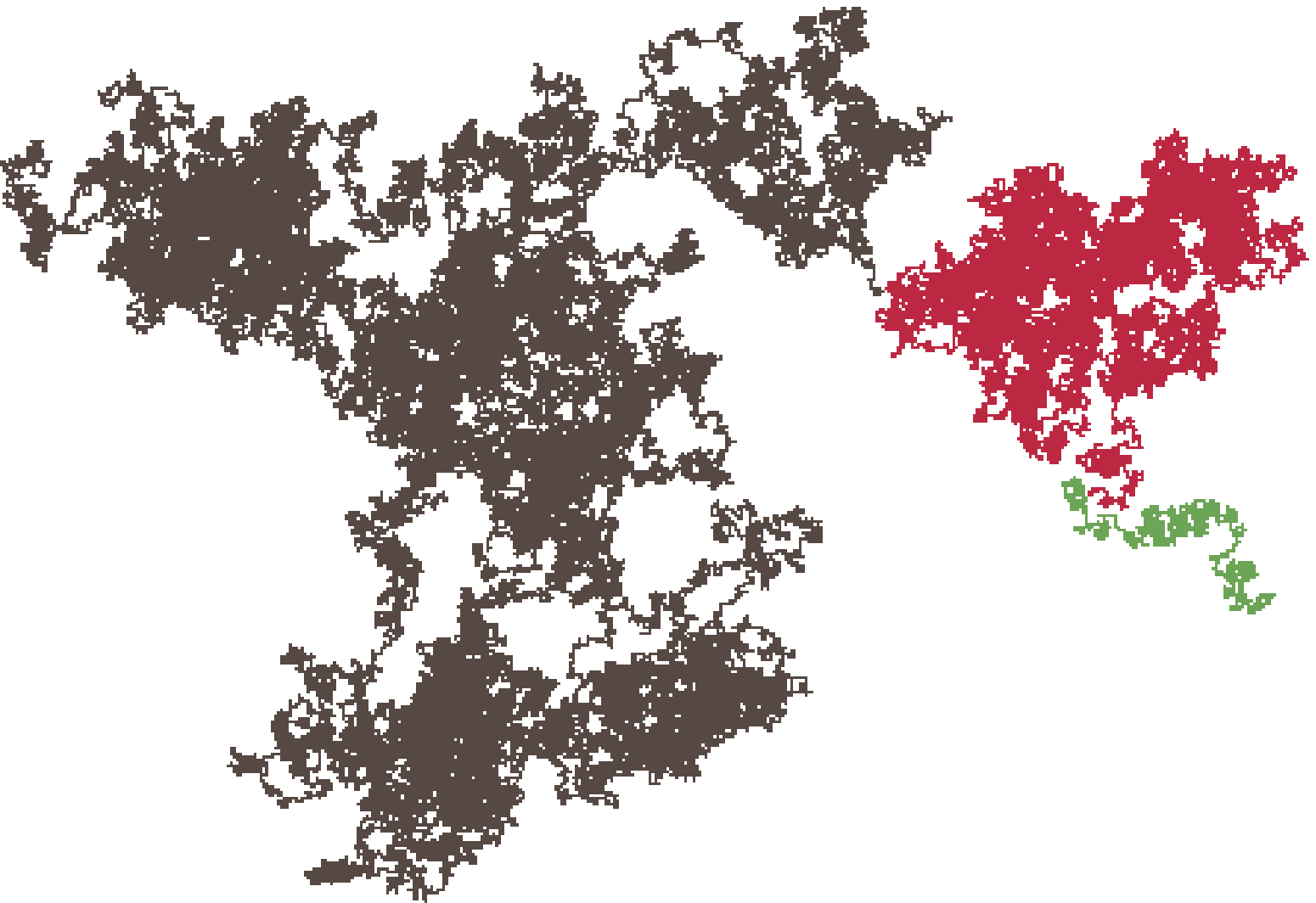}
\narrowtext\caption{\label{Fig2}(Color online) Typical samples of
clusters of the visited sites on the attractive plane by a 3$d$ ARW
of $N=10^6$ shown in different colors, for $\alpha=2$ (left) and
$\alpha=10$ (right). }\end{center}
\end{figure}

Henceforth we investigate the fractal and scaling properties of the
set of all distinct sites visited by the 3$d$ ARW only on the
attractive plane. Each of these sites is visited at least once by
the 3$d$ ARW and marked upon visiting (if not already).\\In this
section, rather than analyzing the  properties of the whole set,
after marking all visited sites on the plane, we identify each
cluster-site as a set of all nearest-neighbor visited-sites on the
lattice with a specific color. Two typical examples of such
clustering are shown in Fig. \ref{Fig2} for two different values of
the strength of attraction $\alpha=2$ and $\alpha=10$. As Fig.
\ref{Fig2} shows, for lower values of $\alpha$, there exist many
isolated clusters of different scales which are accessed by the ARW
only via the third dimension. By increasing the strength of the
attraction, the number of isolated clusters decreases until
$\alpha\rightarrow\infty$ for which, there will be only one large
cluster on the attractive plane.

%%%%%%%%%%%%%%%%%%%%%%%%%%%%%%%%%%%%%%%%%%%%%%%%%%%%%%%%%%%%%%%%%%%%%%%%%%%%%%%%%%%%%%%%%%%%%%%%%%%%%%%%%%%%%%%%
%%%%%%%%%%%%%%%%%%%%%%%%%%%%%%%%%%%%%%%%%%%%%%%%%%%%%%%%%%%%%%%%%%%%%%%%%%%%%%%%%%%%%%%%%%%%%%%%%%%%%%%%%%%%%%%%

To examine possible scale invariance of cluster ensemble for rather
small values of $\alpha$, we compute the cluster size distribution
and check whether it follows a power-law scaling. In the critical
statistical physics, the scaling properties of fractal clusters can
be described by the percolation theory \cite{SA}, where the
asymptotic behavior of cluster distribution $n_s(\lambda)$ near the
critical point $\lambda\rightarrow \lambda_c$ has the following
general form \be\label{Eq1}n_s(\lambda)=
s^{-\tau}F[(\lambda-\lambda_c)s^{\sigma}],\ee where $\sigma$ is an
critical exponent, and the scaling function $F(u)$ approaches to a
constant value for $|u|\ll 1$ and decays rather fast for $|u|\gg 1$.

%%%%%%%%%%%%%%%%%%%%%%%%%%%%%%%%%%%%%%%%%%%%%%%%%%%%%%%%%%%%%%%%%%%%%%%%%%%%%%%%%%%%%%%%%%%%%%%%%%%%%%%%%%%%%%%%
\begin{figure}[t]\begin{center}
\includegraphics[scale=0.4]{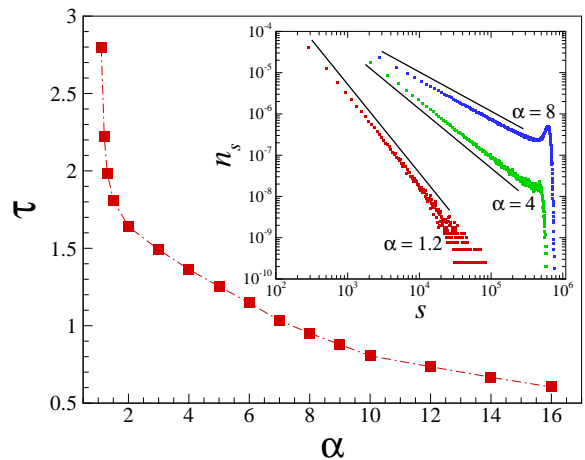}
\narrowtext\caption{\label{Fig3}(Color online) Cluster size
distribution exponent $\tau$ defined in Eq. (\ref{Eq1}), as a
function of the strength of attraction $\alpha$. Inset: number
density $n_s$ of clusters of the visited lattice sites of size $s$
on the attractive plane for three different values $\alpha=1.2$, $4$
and $8$. The solid lines show the power-law behavior in the scaling
region. The error bars are almost the same size as the
symbols.}\end{center}
\end{figure}
%%%%%%%%%%%%%%%%%%%%%%%%%%%%%%%%%%%%%%%%%%%%%%%%%%%%%%%%%%%%%%%%%%%%%%%%%%%%%%%%%%%%%%%%%%%%%%%%%%%%%%%%%%%%%%%%

We undertook simulations for several values of $\alpha$ to measure
the distribution of the cluster sizes of the visited lattice sites
by the 3$d$ ARW on the attractive plane (this is the probability
that a visited lattice site on the attractive plane belongs to a
cluster of size $s$). We gathered ensembles of a number of
$5\times10^4$ (for smaller $\alpha$) and $1.5\times10^6$ (for larger
values of $\alpha$) independent samples of fractal patterns with
marked visited-sites on the attractive plane. The number of the
random walk steps was chosen to be $N=4\times10^6$ in all
simulations. The number density $n_s$ of the clusters of size $s$
has then been computed for each specific value of $\alpha$ by
counting the number of clusters of size $s$ divided by the total
number of all clusters.\\We find that for rather small and
intermediate size scale clusters, the distribution shows a power law
behavior compatible with the scaling relation in Eq. (\ref{Eq1}). As
can be seen in the inset of Fig. \ref{Fig3}, the curves for
different values of $\alpha$ exhibit a sharp drop-off, indicating
indeed that they contain only small clusters. By increasing $\alpha$
the interval for scaling region decreases and a peak appears which
signals the formation of large scale clusters.\\Our estimation of
the cluster size distribution exponent $\tau$ in the scaling region
as a function of $\alpha$ is also shown in Fig. \ref{Fig3}. One
observes that the exponent $\tau$ has a significant dependence on
the strength of attraction $\alpha$.

%%%%%%%%%%%%%%%%%%%%%%%%%%%%%%%%%%%%%%%%%%%%%%%%%%%%%%%%%%%%%%%%%%%%%%%%%%%%%%%%%%%%%%%%%%%%%%%%%%%%%%%%%%%%%%%%
\begin{figure}[t]\begin{center}
\includegraphics[scale=0.39]{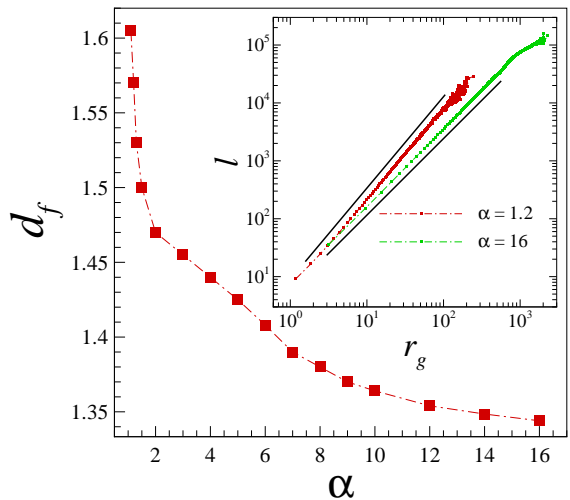}
\narrowtext\caption{\label{Fig4}(Color online) The fractal dimension
of the perimeter of a cluster of visited sites on the attractive
plane by 3$d$ ARW, as a function of the strength of attraction
$\alpha$. Inset: the average length of the perimeter $l$ of a
cluster versus its average radius of gyration $r_g$, for two
different strengths of attraction $\alpha=1.2$ (upper graph) and
$\alpha= 16$ (lower graph). The solid lines show the power-law
behavior in the scaling region. The error bars are almost the same
size as the symbols. }\end{center}
\end{figure}
%%%%%%%%%%%%%%%%%%%%%%%%%%%%%%%%%%%%%%%%%%%%%%%%%%%%%%%%%%%%%%%%%%%%%%%%%%%%%%%%%%%%%%%%%%%%%%%%%%%%%%%%%%%%%%%%

\section{Fractal dimension of the cluster boundaries on the attractive plane}

The remainder of this paper is dedicated to investigate the fractal
properties of the boundaries of the visited-sites clusters on the
attractive plane.\\Given a configuration of visited sites by the
3$d$ ARW on the attractive plane, the first step is to identifying
different clusters as outlined before. After that, the boundary
curve of each isolated cluster has to be identified. However the
definition of interfaces and cluster boundaries on a square lattice
can contain some ambiguities, there has been introduced a
well-defined \emph{tie-breaking} rule in \cite{Saberi} that
generates non-intersecting cluster boundaries on a square lattice
without any ambiguity.\\To define the hull for each identified
cluster according to the algorithm defined in \cite{Saberi}, a
walker (which, of course, has to be distinguished from the 3$d$ ARW)
moves clockwise along the edges of the dual lattice (which is also a
square lattice) around the cluster starting from a given boundary
edge on the cluster. The direction at each step is always chosen
such that walking on the selected edge leaves a visited site on the
right and an empty plaquette on the left of the walker. If there are
two possible ways of proceeding, the preferred direction is that to
the right of the walker. The directions \emph{right} and \emph{left}
are defined locally according to the orientation of the
walker.\\According to this procedure, we have generated an ensemble
of cluster boundary loops for several different strengths of
attraction in the range $1.1\leq\alpha\leq16$. Using the scaling
relation $l\sim r_g^{d_f}$, between the average length of the
perimeter of the loops $l$, and their average radius of gyration
$r_g$, we computed the fractal dimension $d_f$ of the cluster
boundaries as a function of $\alpha$. The results are shown in Fig.
\ref{Fig4}.

The fractal dimension shows again a significant dependence on the
strength of attraction $\alpha$. In the limit
$\alpha\rightarrow\infty$ $d_f$ converges to the value
$\frac{4}{3}=1.3\bar{3}$ which is the fractal dimension of the SAW
i.e., the outer perimeter of the planar Brownian motion.

\section{conclusions}

In this paper, we have studied the scaling properties and the
fractal structure of the visited lattice-sites by a Brownian
particle in 3$d$ which is attracted by a plane with the strength
$\alpha$. The fractal dimensions of the set of visited sites by the
3$d$ random walker in both three dimensions and on the attractive
plane are computed which both converge to the same value $\sim1.83$
for large $\alpha$. We also found that size distribution of the
cluster of visited sites by the particle on the attractive plane,
has a scaling form characterized by an exponent that depends
significantly on the strength of attraction.\\The fractal dimension
of the surrounding loops of the clusters on the plane has been
computed as a function of $\alpha$. This also converges
asymptotically to the expected value for SAW i.e., the external
perimeter of a planar Brownian motion.

These results need however some theoretical framework and
mathematical proof. The other interesting feature which can be
investigated, is the possible conformal invariance of the cluster
boundaries on the attractive plane, which can be treated using SLE
techniques (such study is already done only for the limiting case
$\alpha\rightarrow\infty$ where the problem reduces to a 2$d$ random
walk in the attractive plane whose boundary is described by
SLE$_{8/3}$). The fractal dimension of an SLE$_\kappa$ curve is
given by $d_f=1+\kappa/8$. In case of conformal invariance of
cluster boundaries on the attractive plane in our model, they would
be defined by a diffusivity $\kappa$ which depends on the strength
of attraction.

\textbf{Acknowledgement.} I would like to thank H. Dashti-Naserabadi
for his helps on programming. This work is financially supported by
the National Elite Foundation of Iran, and INSF grant No. 87041917.

\end{document}